\documentclass{elsart5}
 \usepackage{graphicx}

\usepackage{amssymb}
\usepackage{color}
\usepackage{mathrsfs}

\begin{document}

\begin{frontmatter}
% Title, authors and addresses
% use the thanksref command within \title, \author or \address for footnotes;
% use the corauthref command within \author for corresponding author footnotes;
% use the ead command for the email address,
% and the form \ead[url] for the home page:

\title{Magnetic properties of Fe/Dy multilayers : a Monte Carlo investigation}

\author{E. Talbot\corauthref{cor1}}
\ead{etienne.talbot@univ-rouen.fr}
\corauth[cor1]{Tel: + 33-2-32-95-51-32    ; fax: + 33-2-32-95-50-32}
\author{P.E. Berche}
\author{D. Ledue}
\author{R. Patte}
\address{Groupe de Physique des Mat\'{e}riaux, UMR CNRS 6634, Avenue de l'universit\'{e}, BP12, Universit\'{e} de Rouen, F-76801 Saint Etienne du Rouvray Cedex, France}

\begin{abstract}
We investigate the magnetic properties of a Heisenberg ferrimagnetic multilayer by using Monte Carlo simulations. The aim of this work is to study the local structural anisotropy model which is a possible origin of the perpendicular magnetic anisotropy in transition metal/rare earth amorphous multilayers. We have considered a face centered cubic lattice where each site is occupied by a classical Heisenberg spin. We have introduced in our model of amorphous multilayers a small fraction of crystallized Fe-Dy nanoclusters with a mean anisotropy axis along the deposition direction. We show that a competition in the energy terms takes place between the mean uniaxial anisotropy of the Dy atoms in the nanoclusters and the random anisotropy of the Dy atoms in the matrix.
\end{abstract}

%%%%%%%%%use  the \KEY command at the begin of keyword text%%%%%%%%%
\end{frontmatter}

%%%%%%%%%%%%%%%%%%%%%%%%%%%%%%%%%%%%%%%%%%%%%%%%%%%%%%%%%%%%%%%%%%%%%%%%%%%%%%%%%%%%%%%%%%%%%%%%%%%%%%%%%%%%%%%%%%%%%
%%%%%%%%%%%%%%%%%%%%%%%%%%%%%%%%%%%%%%%%%%%%%%%%%%%%%%%%%%%%%%%%%%%%%%%%%%%%%%%%%%%%%%%%%%%%%%%%%%%%%%%%%%%%%%%%%%%%%
%%%%%%%%%%%%%%%%%%%%%%%%%%%%%%%%%%%%%%%%%%%%%%%%%%%%%%%%%%%%%%%%%%%%%%%%%%%%%%%%%%%%%%%%%%%%%%%%%%%%%%%%%%%%%%%%%%%%%
%%%%%%%%%%%%%%%%%%%%%%%%%%%%%%%%%%%%%%%%%%%%%%%%%%%%%%%%%%%%%%%%%%%%%%%%%%%%%%%%%%%%%%%%%%%%%%%%%%%%%%%%%%%%%%%%%%%%%

\section{Introduction}\label{}
Amorphous ferrimagnetic Transition Metal/Rare Earth (TM/RE) multilayers such as Fe/Dy are very attractive since they may exhibit in some conditions interesting magnetic properties: a strong uniaxial perpendicular magnetic anisotropy, a high coercive field and a high Curie temperature for example. The origin of the perpendicular orientation of the magnetisation (relatively to the plane of the substrate) is not yet clearly understood, and different models have been proposed to explain it. These models are based on an anisotropic distribution of TM-RE pairs along the perpendicular direction \cite{SATO}, dipolar interactions \cite{SCHU}, local structural anisotropy \cite{FUJI,MERG} or single-ion anisotropy for compositionally moduled films \cite{SHAN}.

In this paper, we present a Monte Carlo (MC) investigation of a ferrimagnetic multilayer consisting in classical Heisenberg spins with different modulus. In the framework of the local structural anisotropy model, we study the influence of the single-ion anisotropy constant on the RE atoms and of the crystallized nanocluster concentration on the magnetisation orientation of amorphous Fe/Dy multilayers at low temperature.

%%%%%%%%%%%%%%%%%%%%%%%%%%%%%%%%%%%%%%%%%%%%%%%%%%%%%%%%%%%%%%%%%%%%%%%%%%%%%%%%%%%%%%%%%%%%%%%%%%%%%%%%%%%%%%%%%%%%%
%%%%%%%%%%%%%%%%%%%%%%%%%%%%%%%%%%%%%%%%%%%%%%%%%%%%%%%%%%%%%%%%%%%%%%%%%%%%%%%%%%%%%%%%%%%%%%%%%%%%%%%%%%%%%%%%%%%%%
%%%%%%%%%%%%%%%%%%%%%%%%%%%%%%%%%%%%%%%%%%%%%%%%%%%%%%%%%%%%%%%%%%%%%%%%%%%%%%%%%%%%%%%%%%%%%%%%%%%%%%%%%%%%%%%%%%%%%
%%%%%%%%%%%%%%%%%%%%%%%%%%%%%%%%%%%%%%%%%%%%%%%%%%%%%%%%%%%%%%%%%%%%%%%%%%%%%%%%%%%%%%%%%%%%%%%%%%%%%%%%%%%%%%%%%%%%%

\section{Model and simulation technique}
\subsection{Model}
\begin{figure}
\begin{center}
\includegraphics[width=3.5cm]{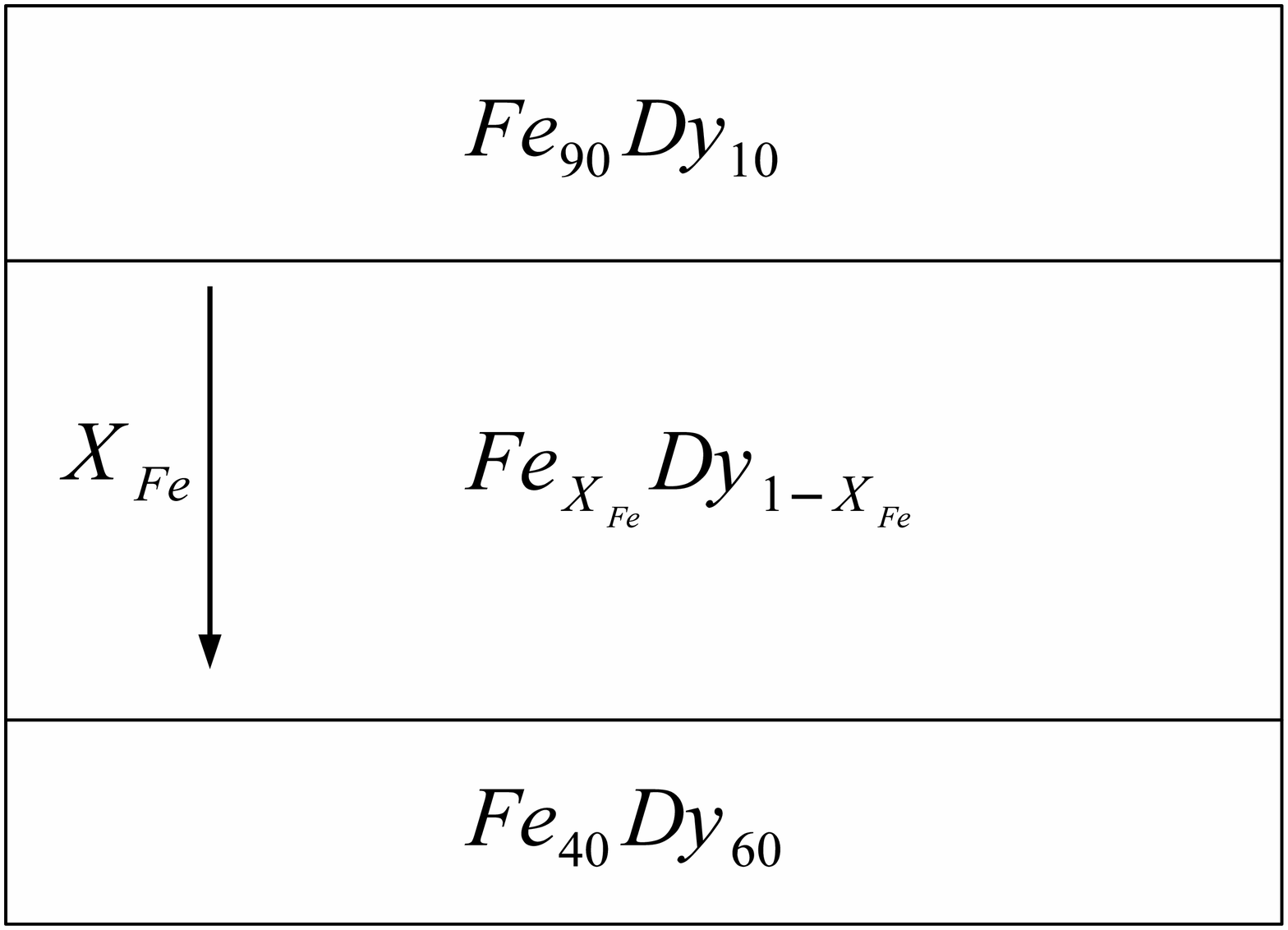}

\caption{Diffusion profile extracted from the experimental investigation of (Fe 3nm/Dy 2nm) multilayers (see ref. \cite{TAMI}).}
\label{profil_sonde}
\end{center}
\end{figure}

We consider a face centered cubic multilayered system made up of Fe and Dy atoms. The Hamiltonian $\mathscr{H}$ of the system is 

\begin{displaymath}
\mathscr{H} = - \sum_{<i,j>} J_{ij} \mathbf{S}_{i} \cdot \mathbf{S}_{j} - \sum_{i \in \rm{Dy}} D_{i} (\mathbf{S}_{i} \cdot \mathbf{n}_{i} )^{2} ,
\end{displaymath}
where $\mathbf{S}_{i}$ is a classical Heisenberg spin, $J_{ij}$ is the nearest-neighbour exchange interaction, $D_{i}$ is the single-ion magnetic anisotropy constant which occurs only for the Dy atoms and $\mathbf{n}_{i}$ is a unit vector along the local easy axis. The following magnetic parameters are those of the free atoms: ${g}_{\rm{Fe}}=2$, ${g}_{\rm{Dy}}=4/3$ and $S_{\rm{Dy}}=5/2$. In order to relate our results to real systems, the $J_{\rm{Fe-Dy}}$ interaction is considered to be negative, whereas $J_{\rm{Fe-Fe}}$ and $J_{\rm{Dy-Dy}}$ are positive. As in previous studies \cite{HEIM}, we consider that $S_{\rm{Fe}}$, $J_{\rm{Fe-Fe}}$ and $J_{\rm{Fe-Dy}}$ are concentration dependent while the minor $J_{\rm{Dy-Dy}}$ interaction is kept constant :
\begin{displaymath}
S_{\rm{Fe}} (X_{\rm{Fe}}) = 1.1 - 1.125 (1 - X_{\rm{Fe}}) ,
\end{displaymath}
\begin{displaymath}
J_{\rm{Fe-Fe}} (X_{\rm{Fe}})/ k_{\rm{B}}  = 14.86 + 86.96(1-X_{\rm{Fe}}) \quad (\textrm{in }  \textrm{ K} ) ,
\end{displaymath}
\begin{displaymath}
J_{\rm{Fe-Dy}}  (X_{\rm{Fe}})/ k_{\rm{B}}  = 1.63 - 38.41 (1-X_{\rm{Fe}}) \quad (\textrm{in } \textrm{ K} ) ,
\end{displaymath}
\begin{displaymath}
J_{\rm{Dy-Dy}}  / k_{\rm{B}}  = 1.16 \quad (\textrm{in }  \textrm{ K} ) ,
\end{displaymath}
where $X_{\rm{Fe}}$ is the Fe concentration in the Fe-Dy alloy. The exchange interactions have been adjusted to obtain pure amorphous Fe and Dy Curie temperatures. In order to describe the amorphous nature of the multilayers, exchange interactions are modulated by a gaussian distribution as a consequence of the distribution of the interatomic distance between nearest neighbours \cite{HAND}. Since the experimental reported values for the single-ion anisotropy constant of Dy may vary with two order of magnitude, this parameter will be considered as free in our simulations. We apply periodic boundary conditions in the plane and free boundary conditions in the deposition direction (arbitrarily chosen as the $z$ axis) to take into account the free surfaces in real systems. The diffusion profile along the perpendicular axis has been obtained from tomographic atom probe analysis on (Fe 3nm/Dy 2nm) multilayers \cite{TAMI} elaborated at 570K which give rise to the larger perpendicular magnetic anisotropy. This profile is characterized by high concentrated Fe layers and $\rm{Dy}_{60}-\rm{Fe}_{40}$ alloy layers with diffuse interfaces between them (Fig. \ref{profil_sonde}).

\subsection{Nanocluster model}
\begin{figure}
 \begin{center}
 \includegraphics[width=3.5cm]{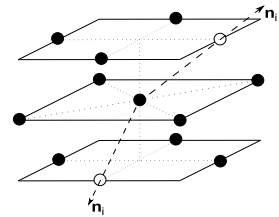}

 \caption{Schematic representation of the nanoclusters with an Fe central atom ($\bullet$) and 2 to 4 Dy atoms ($\circ$) as nearest neighbours. The anisotropy axes of the Dy atoms are represented by the dashed arrows.}
 \label{amasfig}
 \end{center}
 \end{figure}

In order to explain the origin of the perpendicular magnetic anisotropy in Fe/Dy multilayers, we were inspired of the local structural anisotropy model which describes the existence of small crystallized zones with a structure $\rm{RE-TM}_{3}$ or $\rm{RE-TM}_{5}$ \cite{MERG}. This hypothesis has been retained since the experimental investigation of the Fe/Dy multilayers \cite{TAMI,TAMI2} has allowed to reject the other models. We have thus introduced in our model Fe-Dy crystallized nanoclusters made up of 13 atoms (an Fe central atom and its 12 nearest neighbours). We randomly dispatch between 2 and 4 Dy atoms on the neighbouring sites belonging to the upper and lower planes of the central atom (Fig. \ref{amasfig}). Then, the magnetic anisotropy axes of these Dy atoms are along the Fe-Dy bond directions whereas the other Dy atoms display a random anisotropy. We have to mention here that only a few Dy atoms (those of the crystallized nanoclusters) are characterized by anisotropy axes distributed on average along the deposition direction.

\subsection{Numerical simulations}
Numerical simulations were performed by using the importance sampling MC procedure at each temperature based on the standard Metropolis algorithm \cite{METR} based on random trial step \cite{LAND}. Our results were obtained by a slow decrease of the temperature starting from the paramagnetic state.
At each temperature, $5 \times 10^{3}$ Monte Carlo steps (MCS) have been discarded for equilibration before averaging the physical quantities over the following $10^{4}$ MCS. This procedure allows to reach equilibrium at each temperature when the Hamiltonian of the system does not display too much frustration, which is the case here. As a lot of parameters are generated through random procedures (atomic positions and exchange interaction modulation) the thermodynamic quantities have been finally averaged over 20 different configurations in order to get reliable numerical results.

%%%%%%%%%%%%%%%%%%%%%%%%%%%%%%%%%%%%%%%%%%%%%%%%%%%%%%%%%%%%%%%%%%%%%%%%%%%%%%%%%%%%%%%%%%%%%%%%%%%%%%%%%%%%%%%%%%%%%
%%%%%%%%%%%%%%%%%%%%%%%%%%%%%%%%%%%%%%%%%%%%%%%%%%%%%%%%%%%%%%%%%%%%%%%%%%%%%%%%%%%%%%%%%%%%%%%%%%%%%%%%%%%%%%%%%%%%%
%%%%%%%%%%%%%%%%%%%%%%%%%%%%%%%%%%%%%%%%%%%%%%%%%%%%%%%%%%%%%%%%%%%%%%%%%%%%%%%%%%%%%%%%%%%%%%%%%%%%%%%%%%%%%%%%%%%%%
%%%%%%%%%%%%%%%%%%%%%%%%%%%%%%%%%%%%%%%%%%%%%%%%%%%%%%%%%%%%%%%%%%%%%%%%%%%%%%%%%%%%%%%%%%%%%%%%%%%%%%%%%%%%%%%%%%%%%
%%%%%%%%%%%%%%%%%%%%%%%%%%%%%%%%%%%%%%%%%%%%%%%%%%%%%%%%%%%%%%%%%%%%%%%%%%%%%%%%%%%%%%%%%%%%%%%%%%%%%%%%%%%%%%%%%%%%%
\section{Results and discussion}
\begin{figure}
\begin{center}
\includegraphics[width=4.4cm,angle=270]{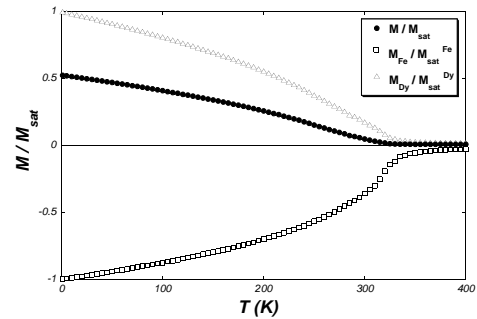}
\caption{Temperature dependence of the reduced magnetisation (global and for each sublattice) of the multilayer without anisotropy. }
\label{Msonde}
\end{center}
\end{figure}

\begin{figure}
\begin{center}
\includegraphics[width=4.4cm,angle=270]{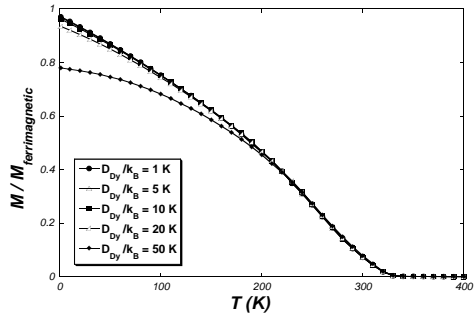}
\caption{Temperature dependence of the total reduced magnetisation with a fraction of $7\%$ Dy atoms embedded in the nanoclusters for different values of the single-ion anisotropy constant $D_{\rm{Dy}}$.}
\label{Mz_Mtot}
\end{center}
\end{figure}

\begin{figure}
\begin{center}
\includegraphics[width=5.8cm]{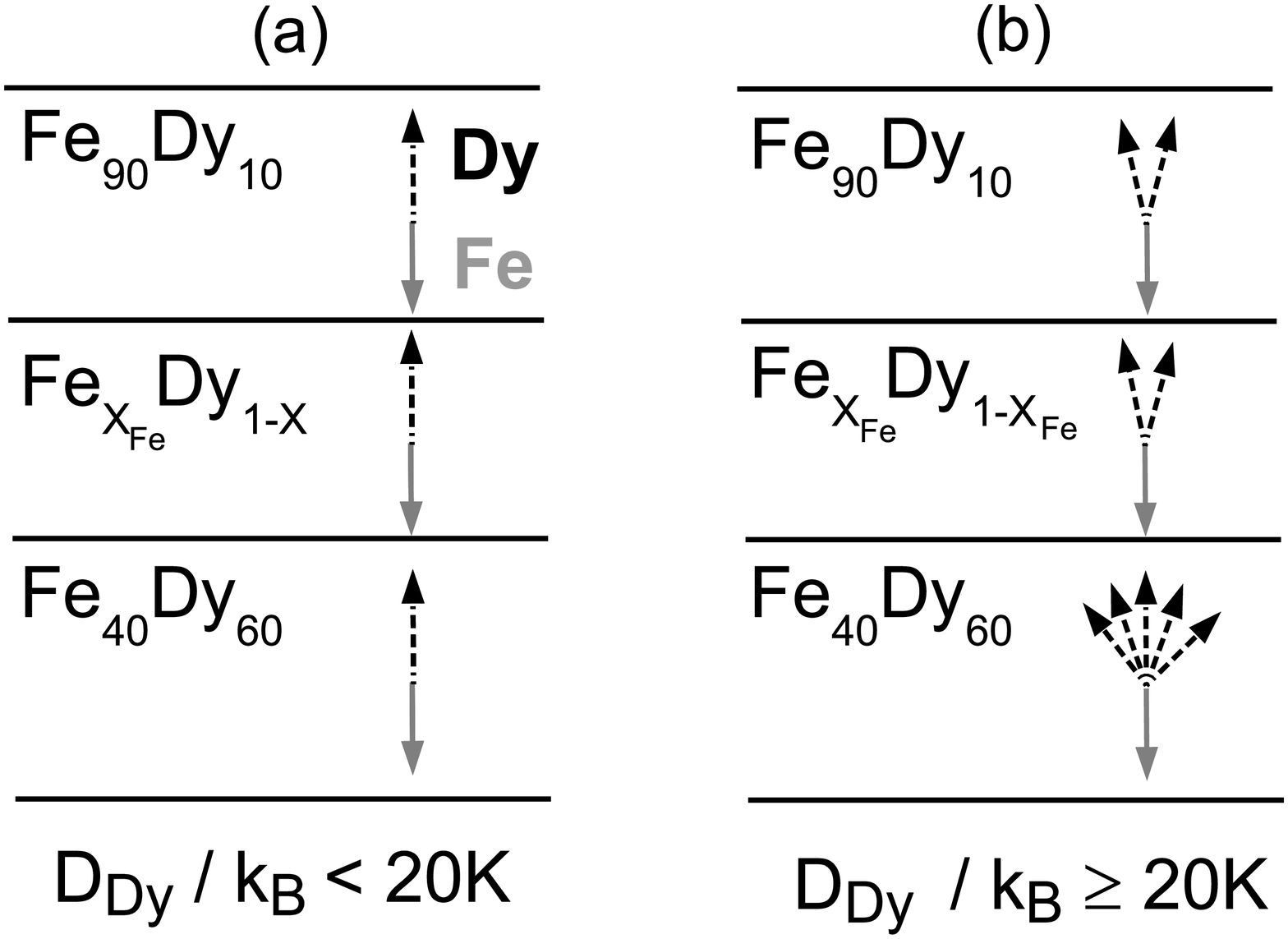}
\caption{Schematic representation of the low temperature magnetic structure when $D_{\rm{Dy}}$ is small (a) or large (b).}
\label{schema}
\end{center}
\end{figure}

As a first step, we have plotted the thermal variation of the total and sublattice magnetisations of the multilayer (consisting in 18 planes of 800 atoms) without any magnetic anisotropy term (Fig. \ref{Msonde}). The multilayer exhibits a single magnetic ordering at about $T_{\rm{C}}=330$K  for the whole sample contrary to the case of multilayers with sharp interfaces between Fe and Dy layers. At low temperature, the total magnetisation of the multilayer is dominated by the Dy sublattice because of its large magnetic moment, and finally the total reduced magnetisation value at 0K is around $0.52$ which corresponds to a collinear ferrimagnetic configuration. In this simulation without magnetic anisotropy, the orientation of the magnetisation does not have any meaning since the Hamiltonian is invariant under any rotation.\\
Then, in the next step, we have studied the influence of the single-ion anisotropy constant $D_{\rm{Dy}}$ on the total magnetisation as a function of the temperature (Fig. \ref{Mz_Mtot}) for the multilayer with $7 \%$ of Dy atoms embedded in the nanoclusters. The total magnetisation is divided by the ferrimagnetic magnetisation since it corresponds to the ground state without anisotropy. We observe that the small values of the single-ion anisotropy constant $D_{{\rm Dy}}$ lead to a strongly ferrimagnetic order at very low temperature with a magnetisation almost perpendicular to the layers as it will be seen later. The larger values of $D_{{\rm Dy}}$ ($D_{{\rm Dy}} / k_{\rm{B}} \ge 20$K) induce a visible decrease of the total magnetisation at low temperature which is the signature of the major influence of the random anisotropy on the Dy atoms not included in the clusters and which gives rise to sperimagnetic structures (Fig. \ref{schema}). Finally, the magnetic configuration is the result of the competition between random magnetic anisotropy on Dy atoms in the matrix, uniaxial anisotropy on average on Dy atoms in the clusters and the exchange interactions.

\begin{figure}
\begin{center}
\includegraphics[width=4.4cm,angle=270]{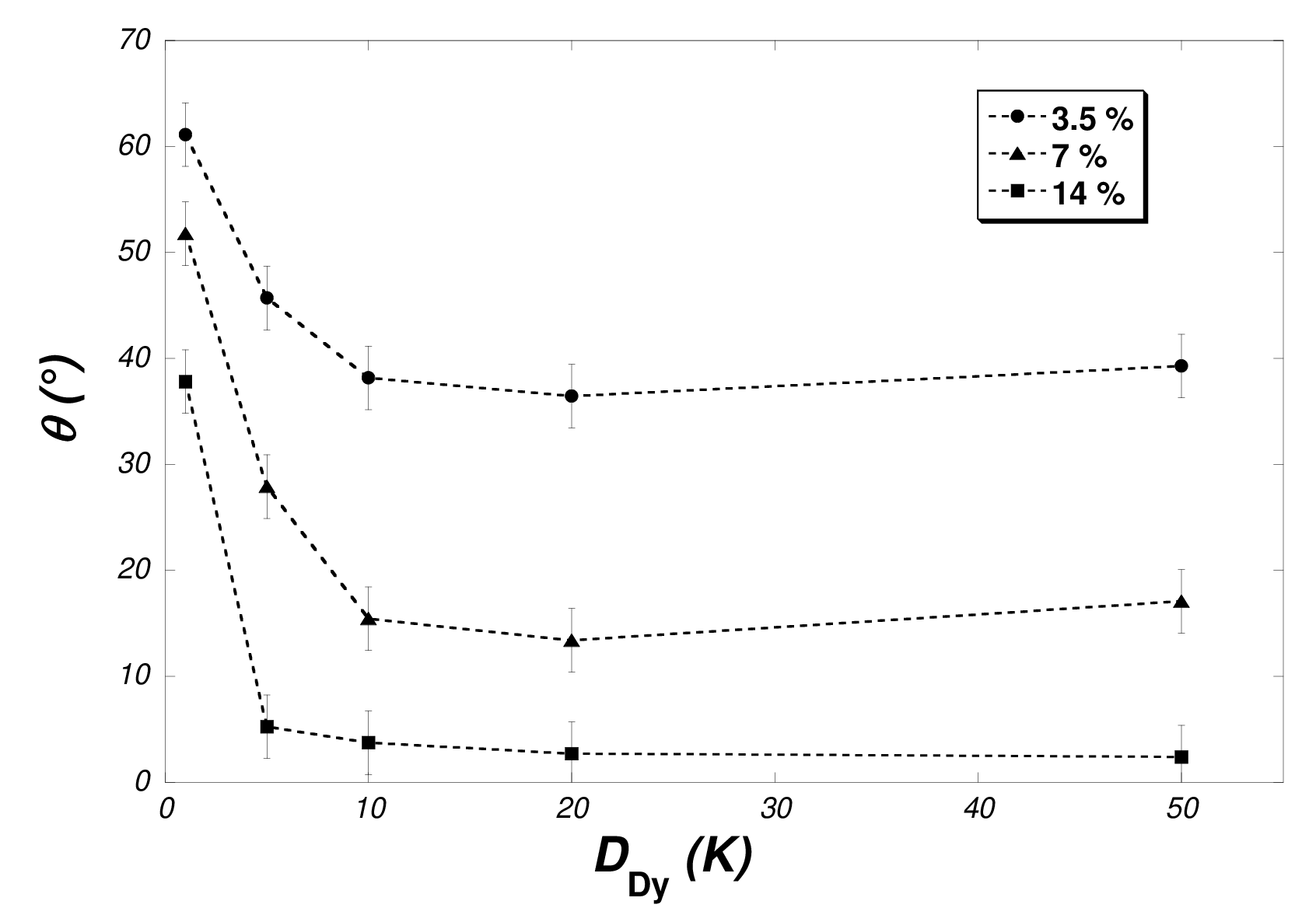}

\vskip -4.1cm
\includegraphics[width=2.cm]{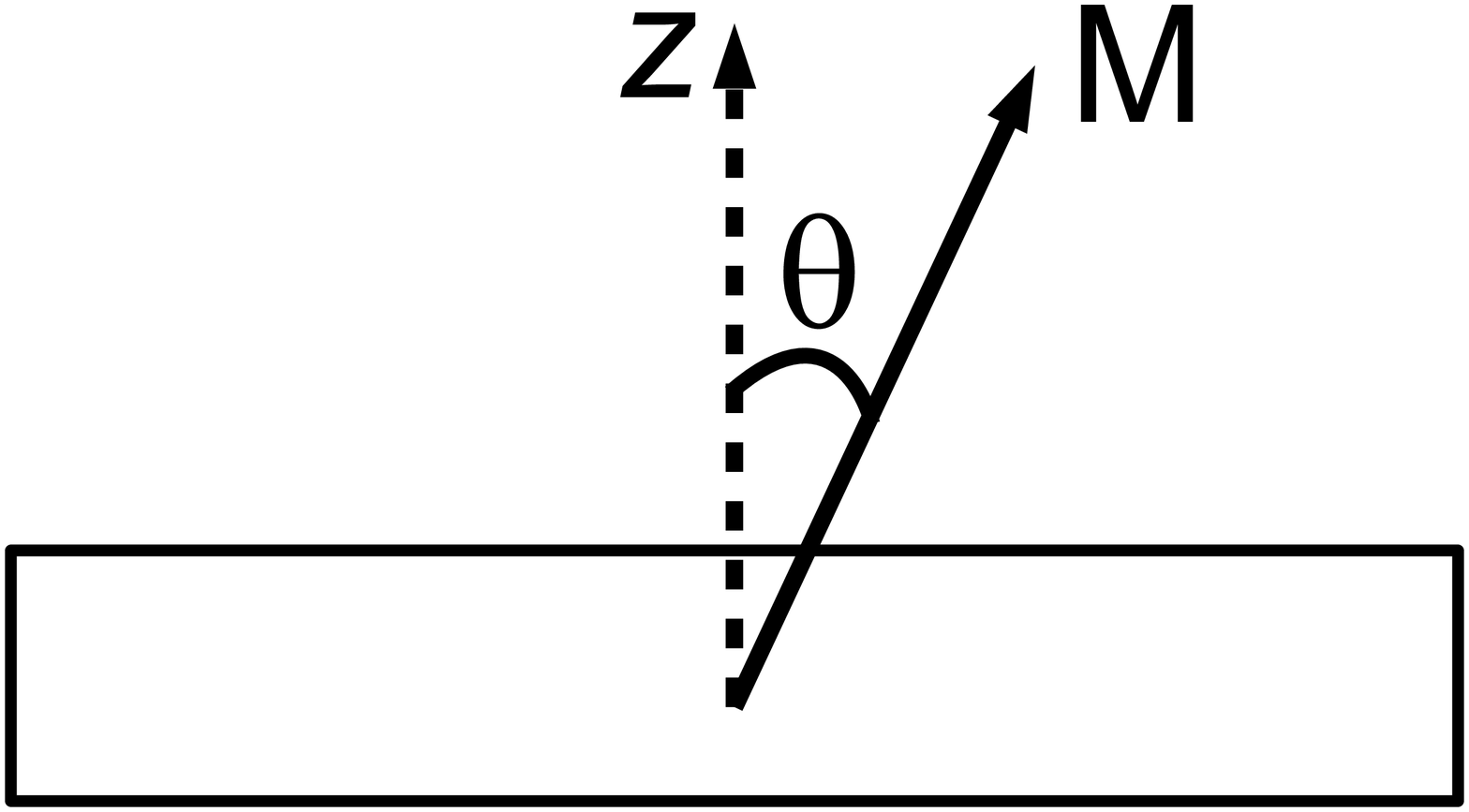}
\vskip 2.5cm
\caption{Angle between the direction of the magnetisation and the perpendicular direction of the multilayer in function of the magnetic anisotropy constant $D_{\rm{Dy}}$ for three values of the fraction of Dy atoms embedded in the crystallized nanoclusters at $T=1$K.}
\label{tetasonde}
\end{center}
\end{figure}

In Fig. \ref{tetasonde}, we have plotted the angle $\theta$ between the direction of the total magnetisation and the $z$-axis in function of the anisotropy constant for three fractions of Dy atoms included in the crystallized nanoclusters ($3.5\%$, $7\%$ and $14\%$). For a given concentration, we observe a decrease in $\theta$ when $D_{\rm{Dy}}$ increases from $1$K to around $20$K which means an increase in the perpendicular component of the magnetisation. For larger $D_{\rm{Dy}}$, we can see a slightly increase in $\theta$ indicating a smaller contribution of $M_{z}$ due to the effect of random magnetic anisotropy of the Dy matrix atoms. We can also notice that the angle $\theta$ is strongly dependent on the nanocluster concentration. A rather small concentration of nanoclusters is sufficient to induce a strong perpendicular magnetisation.

\section{Conclusion}
This work has allowed to investigate the competition between uniaxial and random anisotropy terms in Fe/Dy multilayers as a function of the single-ion anisotropy constant value and of the crystallized nanocluster concentration. The Fe/Dy amorphous multilayer exhibits a major perpendicular magnetisation for only $7 \%$ of Dy atoms in the nanoclusters, and for $D_{\rm{Dy}} / k_{\rm{B}} \ge 10$K. Our results evidence that the crystallized Fe-Dy nanocluster concentration has more influence on the perpendicular magnetic anisotropy than the anisotropy constant value. The simulation of hysteresis loops is actually in progress in order to complete the study of the nanocluster effect on the perpendicular magnetic anisotropy and to get more quantitative results to be compared with experimental data.
\section*{Acknowledgements}
The simulations were performed at the Centre de Ressources Informatiques de Haute Normandie (CRIHAN) under the project No. 2004002.

%\vspace{-3mm}


\begin{thebibliography}{00}
\bibitem{SATO}
N. Sato, J. Appl. Phys. 59 (1985) 2514.
\bibitem{SCHU}
O. Schulte, F. Klose, W. Felsh, Phys. Rev. B 52 (1995) 6480.
\bibitem{FUJI}
H. Fujiwara, X.Y. Yu, S. Tsunashima, S. Iwata, M. Sakurai, K. Suzuki, J. Appl. Phys. 79 (1996) 6270.
\bibitem{MERG}
D. Mergel, H. Heitmann, P. Hansen, Phys. Rev. B 47 (1993) 882.
\bibitem{SHAN}
Z.S. Shan, D.J. Sellmyer, S.S. Jaswal, Y.J. Wang, J.X. Shen, Phys. Rev. B 42 (1990) 10446.
\bibitem{HEIM}
N. Heiman, K. Lee, R. Potter, J. Appl. Phys. 47 (1976) 2634.
\bibitem{HAND}
K. Handrich, Phys. Status Solidi B 32 (1969) K55.
\bibitem{TAMI}
A. Tamion, E. Cadel, C. Bordel, D. Blavette, Scripta Mat. 54 (2006) 671.
\bibitem{TAMI2}
A. Tamion, E. Cadel, C. Bordel, D. Blavette, submitted to Acta Mat. (2006).
\bibitem{METR} 
N. Metropolis, A. Rosenbluth, M. Rosenbluth, A. Teller, E. Teller, J. Chem. Phys. 21 (1953) 1087.
\bibitem{LAND}
D.P. Landau, K. Binder, \textit{A Guide to Monte Carlo Simulations in Statistical Physics}, Cambridge University Press, New York (2000).
\end{thebibliography}
\end{document}